\documentclass[a4paper]{article}
\usepackage{authblk}

\usepackage[english]{babel}
\usepackage[utf8x]{inputenc}
\usepackage[T1]{fontenc}
\usepackage{tikz}
\usepackage{textcomp}

\usepackage[a4paper,top=2cm,bottom=2cm,left=2cm,right=2cm,marginparwidth=1.5cm]{geometry}
\usepackage[none]{hyphenat}   

\usepackage{siunitx}
\usepackage{amsmath}
\usepackage{subfigure}
\usepackage{graphicx}
\usepackage{caption}
\usepackage[colorinlistoftodos]{todonotes}
\usepackage[colorlinks=true, allcolors=blue]{hyperref}

\newcommand*\circled[1]{\tikz[baseline=(char.base)]{
            \node[shape=circle,draw,inner sep=0.5pt] (char) {#1};}}
            
\begin{document}
\date{}

\title{\large{\textbf{Locally resonant metasurfaces for shear waves in granular media}}}

\author[1]{\normalsize{Rachele Zaccherini}}
\author[1]{\normalsize{Andrea Colombi}}
\author[2]{\normalsize{Antonio Palermo}}
\author[1]{\normalsize{Vasilis K. Dertimanis}}
\author[2]{\normalsize{Alessandro Marzani}}
\author[3]{\normalsize{Henrik R. Thomsen}}
\author[1]{\normalsize{Bozidar Stojadinovic}}
\author[1]{\normalsize{Eleni N. Chatzi}}

\affil[1]{\small{\textsl{Department of Civil, Environmental and Geomatic Engineering, ETH Z\"urich, Z\"urich 8093, Switzerland}}}
\affil[2]{\small{\textsl{Department of Civil, Chemical, Environmental and Materials Engineering - DICAM, University of Bologna, Bologna 40136, Italy}}}
\affil[3]{\small{\textsl{Department of Earth Sciences, ETH Z\"urich, Z\"urich 8092, Switzerland}}}

\maketitle

\section*{Abstract}
\begin{sloppypar}
 In this article the physics of horizontally polarized shear waves travelling across a locally resonant metasurface in an unconsolidated granular medium is experimentally and numerically explored. The metasurface is comprised of an arrangement of sub-wavelength horizontal mechanical resonators embedded in silica microbeads. The metasurface supports a frequency-tailorable attenuation zone induced by the translational mode of the resonators. The experimental and numerical findings reveal that the metasurface not only backscatters part of the energy, but also redirects the wavefront underneath the resonators leading to a considerable amplitude attenuation at the surface level, when all the resonators have similar resonant frequency.\\
 A more complex picture emerges when using resonators arranged in a so-called graded design, e.g., with a resonant frequency increasing/decreasing throughout the metasurface. Unlike Love waves propagating in a bi-layer medium, shear waves localized at the surface of our metasurface are not converted into bulk waves. Although a detachment from the surface occurs, the depth-dependent velocity profile of the granular medium prevents the mode-conversion, steering again the horizontally polarized shear waves towards the surface. The outcomes of our experimental and numerical studies allow for understanding the dynamics of wave propagation in resonant metamaterials embedded in vertically inhomogeneous soils and, therefore, are essential for improving the design of engineered devices for ground vibration and seismic wave containment.
\end{sloppypar}

\newpage
\section*{\large{INTRODUCTION}}
Years of study on artificial periodic and locally resonant media for electromagnetic \cite{Pendry1,Pendry2}, acoustic \cite{Ma,Craster} and elastic \cite{Hussein,Craster2} wave control have laid the foundation for the broad application of these structures, known as metamaterials, across several fields and scales. Of particular interest in the context of elastodynamics are metamaterials comprised of subwavelength resonant elements, since they feature spectral bandgaps (i.e., a frequency range of inhibited wave transmission) at low frequencies and, contrary to phononic crystals, they do not rely on periodicity \cite{Liu,Deymier,Liu2}. The ability to mitigate waves and vibrations through the use of the resonance-induced bandgap has been exploited in the context of vibration mitigation in numerous mechanical and civil engineering practical problems, where a surge of metamaterial designs has been proposed \cite{Dalessandro,Brule,Finocchio}. Despite the differences in terms of the size and targeted frequency, the majority of the proposed designs rely on the coupling between a structural component (beam \cite{Zhu,Nouh1} or plate \cite{Xiao,Nouh2,Baravelli}) and an array of mechanical resonators. The coupling occurring between the resonators and the waves propagating inside the component induces a bandgap in the subwavelength regime \cite{Krodel}. In civil engineering for instance, arrays of metre-large mechanical resonators embedded in the soil surface \cite{Vasilis,Achaoui}, typically referred to as elastic metasurfaces, or integrated in the building’s foundation \cite{Salandra,Casablanca} have been recently proposed with the aim of shielding infrastructures from groundborne vibrations and possibly even seismic waves.\\
While shielding is effective inside the bandgap, the corresponding bandwidth is generally too narrow to cover the wide frequency content of incoming waves. Some of the strategies devised to enlarge the bandgap entail the use of non-linear resonators \cite{Lacarbonara} and, as also discussed in this paper, the implementation of a graded design, i.e., an array of resonators with spatially varying resonant frequency \cite{AndreaAlberi}. The theory behind metasurfaces and graded metasurfaces is well established for simple mechanical systems such as plates \cite{AndreaPlate}, waveguides \cite{Maznev,Addouche} and  halfspaces \cite{Colquitt} when Lamb or vertically polarized (e.g. Rayleigh like) waves are considered. Conversely, only a handful of numerical papers have investigated the behavior of metasurfaces interacting with horizontally polarized surface waves. Among these, two studies \cite{PalermoLove,LoveThrees} considered the interaction between Love waves \cite{Achenbach} and mechanical oscillators placed at the free surface of a soft layer overlying a homogeneous elastic half-space. However, actual soils rarely exhibit a bi-layer structure and research considering more complex and realistic media is necessary.\\
In geophysics, a well-established experimental approach uses unconsolidated granular media \cite{Bodet} to mimic an inhomogeneous soil featuring a depth-dependent velocity gradient. The stiffness profile of such media exhibits a power-law dependence on the compacting pressure along depth \cite{Aleshin,Gusev,Makse}. Similar depth dependencies are often adopted also in geotechnical studies \cite{Gazetas,Dakoulas}.\\
In unconsolidated granular media, the inherent gravity-induced stiffness gradient, which causes an upward bending of the rays, combined with the presence of a free surface, enables the propagation of low velocity guided surface acoustic modes (GSAMs). These GSAMs, confined between the surface and the in-depth layer with increasing rigidity, include horizontally (SH) and vertically (P-SV) polarized waves. A recent study \cite{Palermo} investigated the interaction of the P-SV acoustic modes with a metasurface of vertical oscillators. The findings revealed the hybridization of the lowest-order P-SV wave at the metasurface resonance, but an absence of surface-to-bulk mode conversion observed in homogeneous half-spaces, ascribable to the stiffness profile of the granular medium.\\
In this work, we extend the analysis to the dynamics of SH modes, whose existence has been analytically predicted \cite{Aleshin,Gusev}, but is yet to be experimentally observed, offering insights into their interaction with a metasurface of subwavelength horizontal oscillators. By carrying out a small-scale experiment at the IBK Laboratory of ETH Z\"{u}rich, we demonstrate how SH waves propagate through a resonant metasurface embedded in a micrometric granular medium. We then validate the findings by means of dispersion curve analysis and 3D time-transient numerical simulations. To distinguish between Bragg scattering and local resonance phenomena, we measure the SH wavefield within an array of plastic casings buried under the medium surface (here dubbed as "non-resonant casings surface") and compare this with the measurements inside the resonant metasurface. Finally, we numerically and experimentally study the interaction between surface SH waves and a graded metasurface of increasing and decreasing frequency. Our findings allow for understanding the dynamics of wave propagation in resonant metamaterials embedded in soils with an inhomogeneous power-law elastic profile and, therefore, are essential for improving the design of engineered devices for ground vibration and seismic wave containment.

\section*{\large{RESULTS}}
\begin{figure}[ht]
\centering
\includegraphics[width=1\textwidth]
{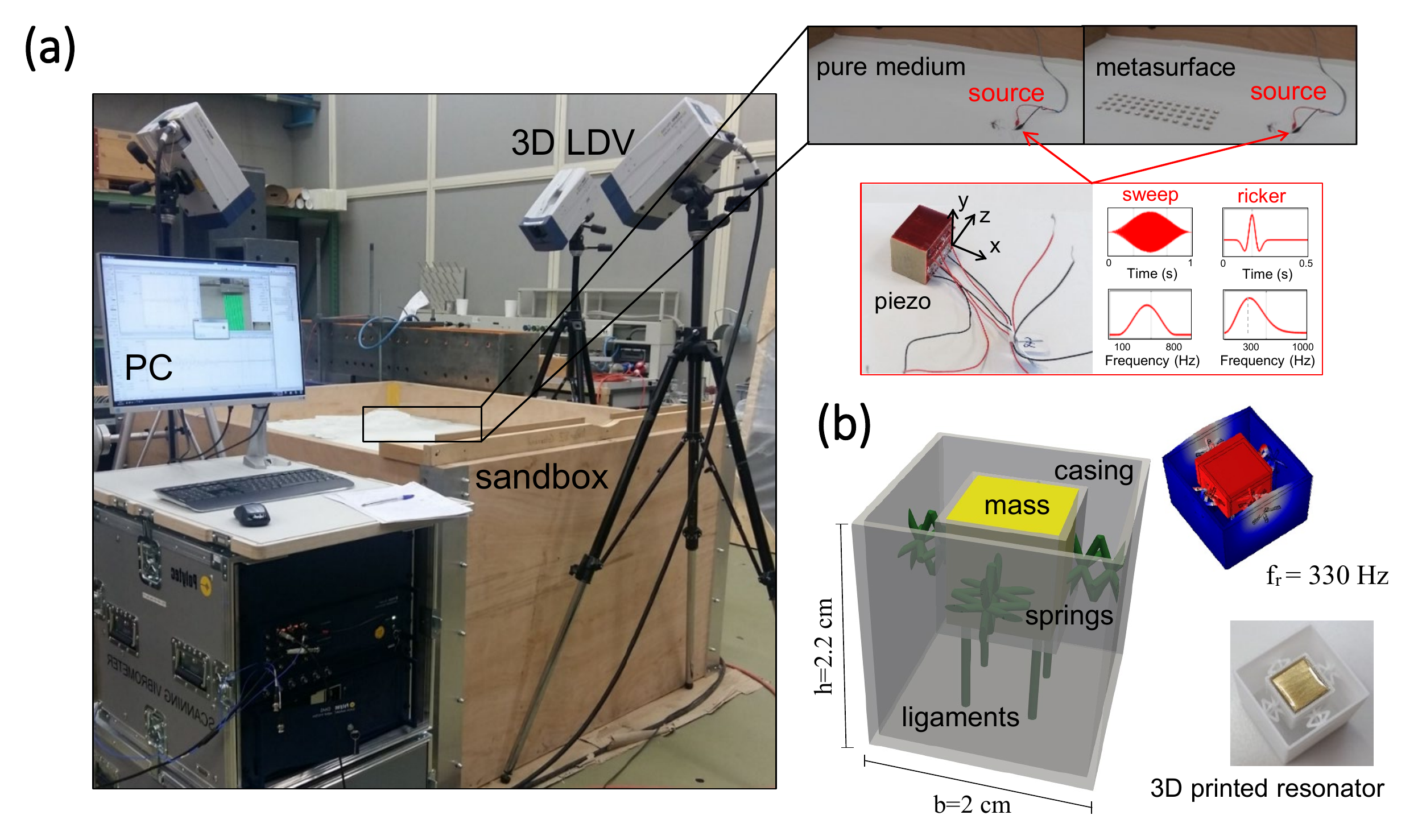}
\caption{\label{fig:f1} Experimental setup and mechanical resonators. (a) Experimental setup involving a wooden box filled with granular material, a 3D laser Doppler vibrometer, an array of subwavelength horizontal oscillators and a 3-axial piezoelectric actuator. (b) Model of the elementary unit cell including a squared brass mass, four truss-like springs, four ligaments acting as vertical supports and an outer casing. First translational mode of the resonant unit together with a picture of the assembled 3D printed resonator.}
\end{figure}

\begin{figure}[ht]
\centering
\includegraphics[width=1\textwidth]
{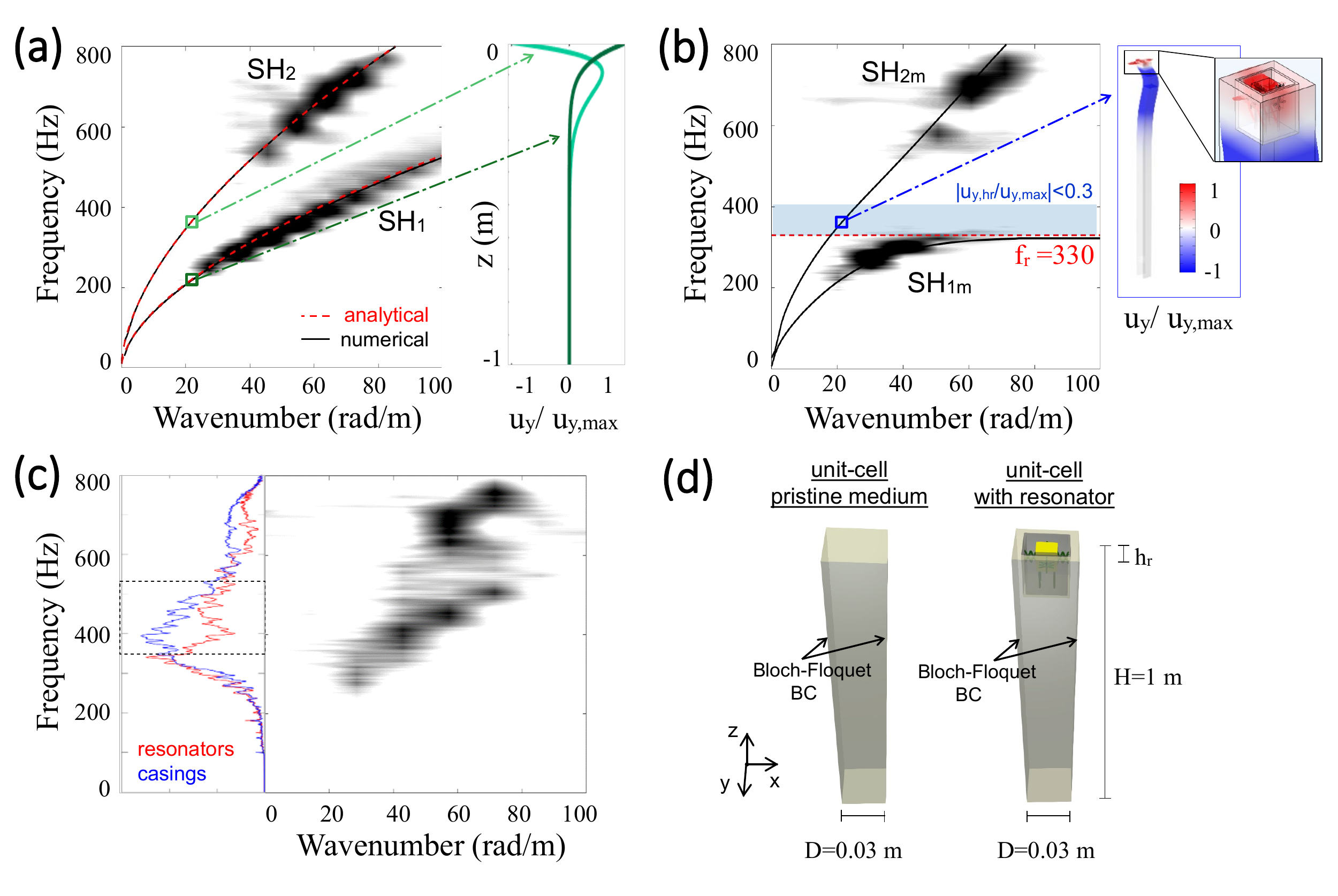}
\caption{\label{fig:f2} Dispersion analysis. (a) Experimental, numerical (black line) and analytical (red dashed line) dispersion curves of the first two low-order SH modes localized at the surface of the granular medium together with the corresponding shape modes for a value of $k=20$ rad/m. (b) Experimental and numerical (black line) dispersion curves of the first two low-order SH modes propagating through the resonant metasurface together with the shape mode of $SH_{2m}$ for a value of $k=20$ rad/m. (c) Experimental dispersion curves of shear horizontal waves propagating through the non-resonant casings-surface. The lateral inset shows the average particle velocity measured in the frequency domain after one line of resonators (red line), and after one line of casings (blue line). (d) Drawing of the 3D unit cells with and without the resonators developed in COMSOL Multiphysics.}
\end{figure}

\begin{figure}[ht]
\centering
\includegraphics[width=1\textwidth]
{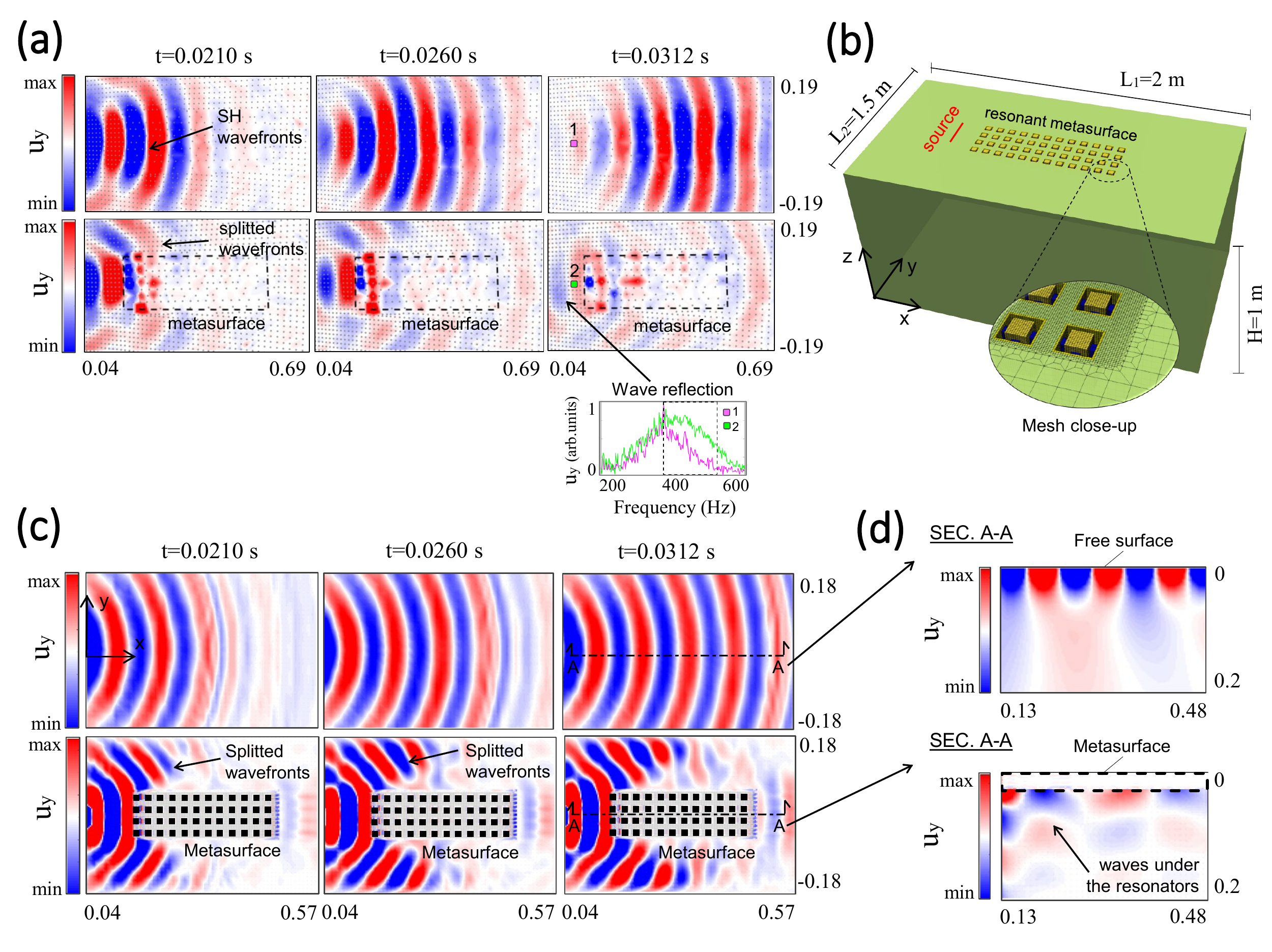}
\caption{\label{fig:area} Experimental and numerical displacement field with and without the metasurface. (a) Experimental displacement field of shear horizontal waves propagating at the free surface of the granular medium (top snapshots) and through the resonant metasurface (bottom snapshots) at three sequential time instants ($t_1=0.0210$ s, $t_2=0.0260$ s, $t_3=0.0312$ s). (b) Computational domain characterized by the granular substrate and the resonant metasurface excited by a harmonic line source at 400 Hz. The inset shows the adaptive mesh close-up. (c) Numerical displacement field of shear horizontal waves propagating at the free surface of the granular medium (top snapshots) and through the resonant metasurface (bottom snapshots) at three sequential time instants ($t_1=0.0210$ s, $t_2=0.0260$ s, $t_3=0.0312$ s). (d) Vertical cross-sections through the centre of the model showing the numerical displacement field of shear horizontal waves traveling at the free surface (top snapshots) and through the metasurface (bottom snapshots) at $t=0.0312$ s.}
\end{figure}

\begin{figure}[ht]
\centering
\includegraphics[width=1\textwidth]
{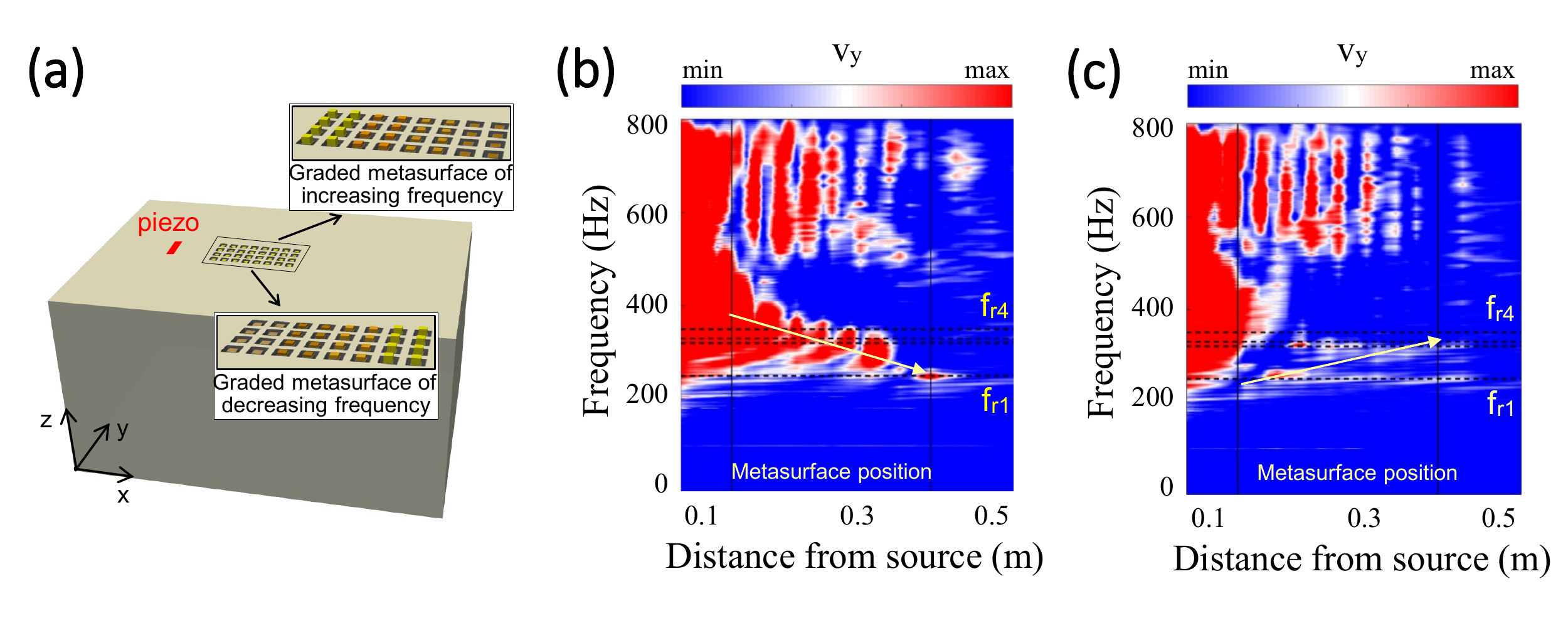}
\caption{\label{fig:f3} Experimental velocity field within graded metasurfaces. (a) Outline of the experimental domain with resonant metasurfaces of gradually increasing and decreasing frequency. (b) Experimental particle velocity as a function of the frequency and the distance from the source for the case of graded metasurface of decreasing frequency. (c) Same as (b) but for a graded metasurface of increasing frequency.}
\end{figure}

\begin{figure}[ht]
\centering
\includegraphics[width=1\textwidth]
{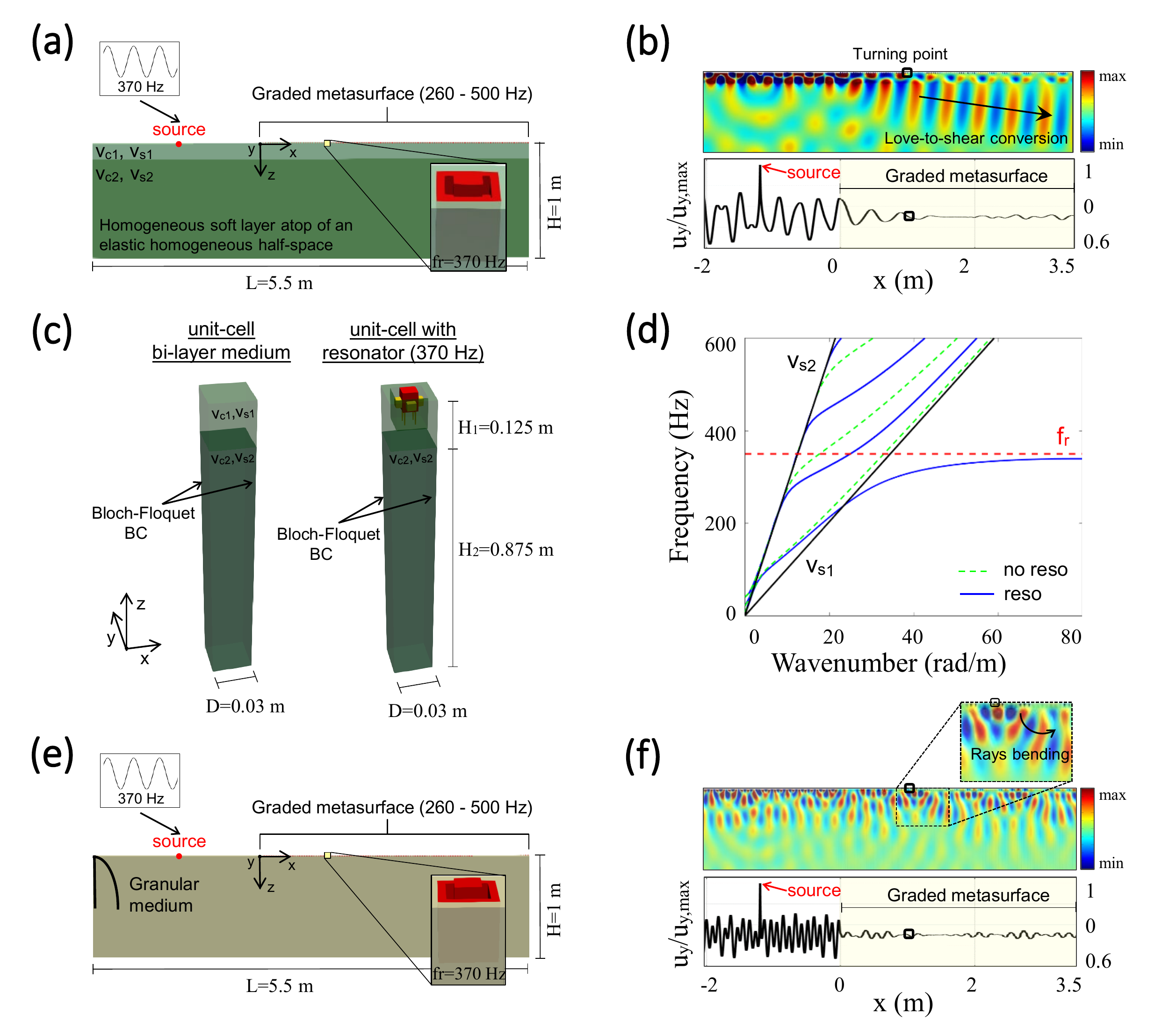}
\caption{\label{fig:f4} Frequency domain numerical simulations. (a) Numerical model developed in COMSOL Multiphysics to simulate the interaction between a graded metasurface of increasing frequency and Love waves. (b) Displacement field of harmonic Love waves traveling at 370 Hz through the graded metasurface and normalised shear displacement at the surface. (c) Drawing of the 3D unit cells with and without the resonators in case of a bi-layer medium. (d) Numerical dispersion curves of Love waves propagating in a bi-layer medium (green dashed curves) and interacting with a metasurface of horizontal resonators (blue curves). The red dashed line marks the resonance frequency of the metasurface, whereas the black lines represent the dispersion curves of shear bulk waves in the upper layer and in the substrate, respectively. (e) Numerical model developed in COMSOL Multiphysics to simulate the interaction between a graded metasurface of increasing frequency and shear horizontal waves propagating in a granular medium. (f) Displacement field of harmonic shear horizontal waves traveling at 370 Hz through the graded metasurface and normalised shear displacement at the surface.}
\end{figure}

\begin{figure}[ht]
\centering
\includegraphics[width=1\textwidth]
{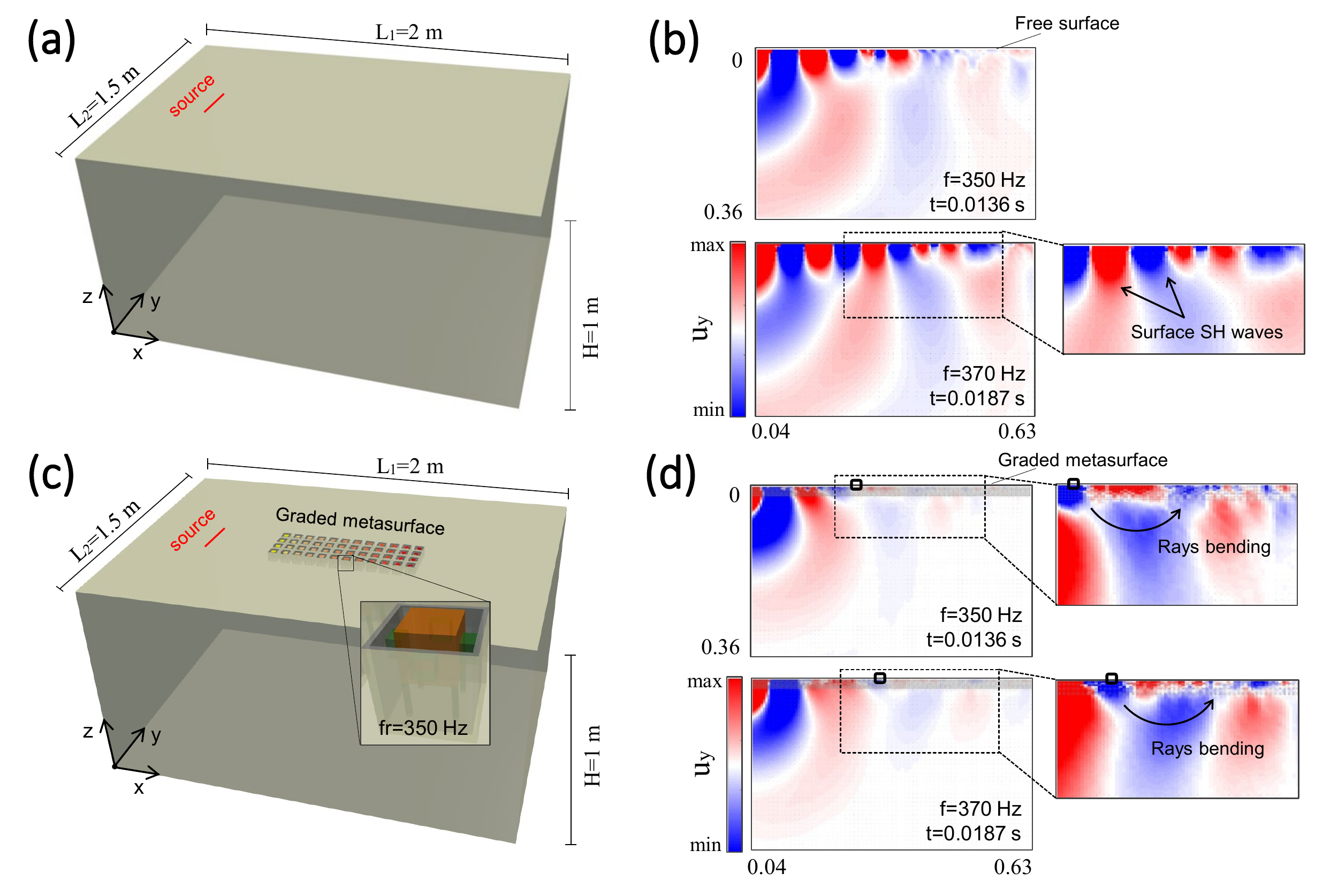}
\caption{\label{fig:f5} Full-scale time domain numerical simulations. (a) Computational domain modeled for the reference case characterized by the granular substrate excited by a harmonic line source (350 and 370 Hz). (b) Displacement field of harmonic shear horizontal waves traveling at the free surface of the granular substrate at two different excitation frequencies (350 and 370 Hz). (c) Computational domain characterized by the granular substrate and the graded metasurface excited by a harmonic line source (350 and 370 Hz). (d) Displacement field of harmonic shear horizontal waves traveling through the graded metasurface at two different excitation frequencies (350 and 370 Hz).}
\end{figure}

\textbf{Experimental setup and data acquisition}. The scaled experimental setup shown in Fig. \ref{fig:f1}(a) includes a wooden box (2000 x 1500 x 1000 mm) filled with granular material (150 µm-diameter glass beads with density of 1600 Kg/m$^3$), a Polytech PSV-500 3D laser Doppler vibrometer, capable of recording the velocity field along three normal directions, and a Physik Instrumente 3-axial piezoelectric actuator driven by a PiezoDrive amplifier. The dimensions of the box are chosen to minimise the interaction between the primary signal with reflections generated by lateral and bottom surfaces within the time window of interest. In addition, to mitigate the amplitude of the reflected signals, a paperboard layer is placed at the base of the box. Similar experimental setups have been used to study the propagation of vertically polarized seismic-waves in unconsolidated and porous media \cite{Bodet} and to investigate their interaction with a metasurface of vertical oscillators \cite{Palermo}. The granular material is gently poured and leveled to ensure a uniform compaction and the realization of a planar surface. The piezoelectric actuator, fixed to a steel block (150 x 100 x 40 mm) buried at 20 mm depth, excites the granular material generating elastic waves polarized in the horizontal plane (SH waves) and localized near the surface. To do so, we only activate the piezoelectric stack corresponding to the y-axis (Fig. \ref{fig:f1}(a)). As input signals (Fig. \ref{fig:f1}(a)), we employ both a modulated chirp from 100 to 800 Hz, useful to derive the dispersion properties and a Ricker wavelet centered at 300 Hz to realize a pulse-like excitation. To limit the effect of non-linearities between the actuator and the granular material, the maximum displacement excursion of the actuator is kept below 2 $\mu$m.\\
The metasurface is built by embedding 48 subwavelength mechanical oscillators at the surface of the granular medium. Each resonator consists of a squared mass supported by four ligaments, which prevent vertical motion, four horizontal truss-like springs and an external casing (height 22 mm, width 20 mm, see Fig. \ref{fig:f1}(b)). The resonator casing and connectors are 3D printed in versatile plastic ($E=1700$ MPa, $\nu=0.3$, $\rho=450$ Kg/m$^3$) using the selective laser sintering technique, which allows the fabrication of the horizontal springs with no need for material support. The resonating mass is made out of brass ($E=110$ GPa, $\nu=0.33$, $\rho=9602$ Kg/m$^3$). The resonant frequency of the mechanical oscillators can be tailored by adjustment of the inclination angle and length of the beams forming the truss-like springs. By design, we set the horizontal resonant frequency of the generic resonator, accounting for the coupling with the surrounding granular material, at $f_{r}=330$ Hz. This resonance value is chosen after experimentally identifying the frequency content of the SH modes propagating in the pristine granular medium, in order to observe potential hybridization phenomena in a frequency range where the waves are highly excited. We experimentally validate this resonance by exciting the oscillators with the piezoelectric actuator. The metasurface is then obtained by arranging the resonators on the medium surface in a 4 x 12 rectangular grid (see Fig. \ref{fig:f1}(a)).\\
For the graded metasurfaces, oscillators with different masses are fabricated. In particular, we assemble four different types of resonators ($f_{r1}=250$ Hz, $f_{r2}=320$ Hz, $f_{r3}=330$ Hz, $f_{r4}=360$ Hz) and arrange them according to increasing and decreasing frequencies. In addition, a non-resonant surface is realized by embedding an identical 4 by 12 arrangement of resonator casings, hosting an internal brass mass with no springs. For all adopted configurations, i.e., pristine granular surface, metasurfaces and non-resonant casings, we use the 3D vibrometer to scan the surface velocity of the granular medium both along the symmetry-axis of the box with a constant step of $\Delta$x=7 mm and over a 380x650 mm area with constant spatial steps of $\Delta$x=12 mm. For each acquisition point we average 10 signals of 0.5 s duration at a sampling rate of 5 kHz.\\
\\
\textbf{Dispersion analysis of the metasurface}. The 2D Discrete Fourier Transform (DFT) of the velocity data recorded along the central horizontal line and representing the dispersion curves are depicted in Figs. \ref{fig:f2}(a)-(b). For the reference case (i.e., the pristine granular material without resonators) the fundamental ($SH_1$) and the $2^{nd}$ higher order ($SH_2$) mode can be clearly identified (Fig. \ref{fig:f2}(a)). While the fundamental mode dominates the spectrum from 200 to 500 Hz with lower velocities (∼ 60 m/s), $SH_{2}$ prevails at higher frequencies, traveling at higher velocities (∼ 80 m/s). Figure \ref{fig:f2}(b) shows the dispersion curves extracted within the resonant metasurface. The $SH_1$ mode interacts with the collective resonances of the metasurface yielding a classical hybridization mechanism between the propagating surface wave and the standing mode of the metasurface. This coupling results in a flat, slow mode (labelled here "$SH_{1m}$" where "$m$" stands for metasurface), which converges asymptotically to the oscillator resonance frequency. Indeed, the dynamics of this $SH_{1m}$ mode resembles the one observed for the Love fundamental mode propagating in a bi-layered medium and interacting with a metasurface of horizontal oscillators \cite{PalermoLove}. Since the resonator spacing is more than 6 times larger than the smallest wavelength of SH waves at the resonance $f_r$=330 Hz, Bragg scattering should not play a role. This is confirmed in Fig.~\ref{fig:f2}(c), where the 2D DFT of the velocity data is computed for the case where the resonators are replaced with the non-resonant casings, equipped with internal masses equivalent to those of the resonators. The two SH modes are similar to the $SH_{1}$ and $SH_{2}$ observed in the reference case of Fig. \ref{fig:f2}(a). No hybridization and no bandgap occur. The inset at left depicts the average particle velocity measured in the frequency domain after one line of resonators (red line), and after one line of casings (blue line). Prior to reaching the resonant frequency of the oscillators (330 Hz) the two signals are similar. After 330 Hz the particle velocity within the metasurface presents a dip inside the bandgap frequency range, centered around 400 Hz. This result confirms that the attenuation zone detected in Fig.~\ref{fig:f2}(b) can be ascribed to the resonant behaviour of the metasurface.\\
We now numerically investigate the dispersive properties of SH waves traveling both at the surface of the granular material and within the metasurface using a Bloch wave \cite{Bloch}–finite element (FE) approach. In particular, the dispersive analysis is numerically carried out by modeling a 3D FE model of the unit cell with and without the resonator (see Fig. \ref{fig:f2}(d)). To obtain the Bloch form wave solution, the velocity profile of the granular medium along the depth is required. Since closed form solutions of the eigenvalue problem are available for SH waves in granular media \cite{Aleshin}, we use the analytical equation of the dispersion relation to estimate the velocity wave profile. This solution has been obtained by studying the dynamics of SH waves traveling in granular media using the GSAM theory described in \cite{Aleshin}, where the granular medium is considered as a linear elastic continuum. The theoretical compressional $v_{c}$ and shear $v_{s}$ velocity profiles of the granular medium are modelled with a power-law dependency on the compacting pressure $p$ along the depth $z$: $v_{c,s}=\gamma_{c,s}(p)^{\LARGE{\alpha_{c,s}}}$=$\gamma_{c,s}(\rho gz)^{\LARGE{\alpha_{c,s}}}$, where $\rho$ is the medium density, $g$ the gravitational constant and $\gamma_{c,s}$, $\alpha_{c,s}$ represent the power-law parameters.\\
In the GSAM framework, the scalar displacement field associated with the SH modes, localized at the surface, $u_y=u_y(z)$ can be described via a depth-dependent Helmholtz equation:
\begin{equation}
 ({v_s}^2(z)u'_y)'+\Big((f/2\pi k)^2-{v_s}^2(z)\Big)u_y=0,
 \label{eq:Helmholtz}
\end{equation}
where $k$ and $f$ are the wavenumber and the frequency, respectively.
The governing equation is complemented with the following boundary conditions:
\begin{equation}
 {v_s}^2(z)u'_y(z))|_{z=0}=0,
   \quad\quad
 u_y(z))|_{z\rightarrow \infty}=0,
 \label{eq:BC1}
\end{equation}
where the first one corresponds to the absence of shear stress $\sigma_{zy}$ at the surface and the second one ensures that waves vanish towards depth. Plugging Eq. \ref{eq:BC1} into \ref{eq:Helmholtz} yields the SH wave characteristic equation, which is analytically solved by applying the geometrical acoustic method \cite{Aleshin}. The resulting eigenvalues are
\begin{equation}
 {\Omega_n}^2=\bigg\{\pi \bigg[ 2n-1- \frac{2\alpha_s-1}{2\alpha_s-2} \bigg] \alpha_s \frac{\Gamma((1/\alpha_s+2)/2)}{\Gamma((1/\alpha_s-1)/2)\Gamma(3/2)} \bigg\}^{2\alpha_s},
 \label{eq:omega}
\end{equation}
where $\Gamma$ represents the Gamma function and the integer $n$ is the $n$-th order mode.\\
Once the eigenvalues are obtained, the dispersion curve of the $n$-th shear horizontal mode can be analytically expressed as follows:
\begin{equation}
 f=\frac{\Omega_n\gamma_{s}(\rho g)^{\LARGE{\alpha_s}}k^{1-\alpha_s}}{2\pi}.
 \label{eq:anadisp}
\end{equation}
By fitting the experimental dispersion curve of the fundamental SH wave ($SH_{1}$) using Eqs. \ref{eq:omega} and \ref{eq:anadisp}, the power-law parameters for the shear profile are derived as $\alpha_s=0.42$, $\gamma_s=3.82$. The compressional velocity profile, which does not affect the SH wave propagation (see Eq. \ref{eq:Helmholtz}) nor the resonator, is derived assuming a Poisson ratio of $\nu=0.376$ as in \cite{Aleshin}.\\
The velocity profiles are then inserted in the 3D unit cell FE models with and without the resonator, developed in COMSOL Multiphysics$^{\circled{\tiny{R}}}$. The 3D granular unit cell is 1 m high ($H$=1 m) and 0.03 m x 0.03 m wide ($D$=0.03 m). The resonator model embedded in the granular material features the same geometric and material properties (mass, stiffness and resonant frequency) of the 3D printed mechanical oscillators. The medium, modeled as a linear elastic continuum, is defined through its density (constant) and velocity profiles (depth-dependent). For both the reference and the metasurface case, absorbing conditions are applied at the bottom edge and periodic Bloch-Floquet boundary conditions on the lateral faces. We perform an Eigenfrequency analysis by sweeping the wave vector in the range $0-k_{max}$ (with $k_{max}=\pi/D$=105 rad/m) to obtain the numerical dispersion relations. We observe an infinite number of SH modes with increasing phase velocity and, consequently, increasingly deep penetration into the medium, as further confirmed in recent studies \cite{Aleshin,Gusev}. Figure \ref{fig:f2}(a) depicts the numerical dispersion curves of the first two SH modes (black lines), which overlap with the analytical ones (red dashed lines), as derived from Eq. \ref{eq:anadisp}, for the reference case. These numerical dispersion relations further lie in good agreement with the experimental curves. We stress that this is the first experimental confirmation of the GSAM prediction for SH waves. The inset in Fig. \ref{fig:f2}(a) illustrates the mode shapes for $SH_1$ and $SH_2$ for a value of $k$=20 rad/m. The maximum displacement occurs at the surface consistently to any surface mode. Moreover, the number of “phases” in the shear displacement profile as a function of depth corresponds to the order of modes, as revealed in \cite{Aleshin,Gusev}. Figure \ref{fig:f2}(b) shows the numerical dispersion curves of the first two SH modes interacting with the resonant metasurface (black lines). The hybridization between the local resonances and the fundamental mode is confirmed by the numerical analysis. Although a frequency gap does not occur, meaning that higher order SH waves do not hybridize with the mechanical oscillators, the particle displacement of the surface layer enclosing the resonators is close to zero. The maximum horizontal displacement is now localized under the resonator (see the inset in Fig. \ref{fig:f2}(b)), suggesting a possible SH wave propagation under the resonant metasurface. The shaded light-blue area of Fig. \ref{fig:f2}(b) identifies the frequency range where the surface displacement is negligible. This is defined as $u_{y,h_r}/u_{y,max}<0.3$, where $u_{y,h_r}$ is the shear horizontal displacement calculated at the resonator embedding depth $h_r$ (i.e., at the height of the centre of gravity of the resonant mass, see Fig. \ref{fig:f2}(d)) and $u_{y,max}$ the maximum displacement along the depth. The existence of this frequency zone of quasi-zero displacement detected in proximity of the resonant frequency of the metasurface explains the energy gap observed in the experimental dispersion relation obtained from surface velocity signals (Fig. \ref{fig:f2}(b)).\\
\\
\textbf{Wavefield analysis in the metasurface}. 
The experimental displacement fields of SH waves propagating at the free surface and within the resonant metasurface filtered inside the bandgap (330-500 Hz) are compared at three sequential time instants ($t_1$=0.0210 s, $t_2$=0.0260 s, $t_3$=0.0312 s) in Fig.~\ref{fig:area}(a). For the reference case without the resonators (top snapshots) nearly plane SH wavefronts propagate in the granular material. Conversely, when traveling within the metasurface (bottom snapshots), the SH wavefronts impinging the mechanical oscillators are scattered and attenuated. 
The inset focuses on the displacement spectrum of a point on the surface located before the first resonator line in the reference (point 1, purple line) and the metasurface (point 2, green line) case. The comparison between the two signals reveals the frequency content of the reflected signal, which mainly falls inside the bandgap. Although of small intensity, reflected wavefronts due to local resonances are visible in front of the metasurface in the third bottom snapshot at $t_3$=0.0312 s.\\
To verify these experimental observations and analyse the anatomy of the SH wavefield interacting with the resonators, we perform full-scale 3D numerical simulations with and without the metasurface using SPECFEM3D, a parallel code based on the spectral element method, able to solve large-scale time dependent elastodynamic problems. The computational domain, depicted in Fig. \ref{fig:area}(b), is a 1 m deep half-space of granular medium, 2 × 1.5 m wide, modeled as a linear elastic continuum described by the previously estimated power-law velocity profile. While the upper surface of the domain is traction free to support surface wave propagation, absorbing conditions are applied to the other boundaries to avoid reflections from the bottom and side edges of the box. A line source, applied at a distance of 0.2 m from the left boundary, is used to generate SH waves. The time history of the excitation is described by a sine wave at 400 Hz, a frequency value which falls within the experimental bandgap. The metasurface comprises 48 resonators arranged in a 4 x 12 rectangular grid and featuring the properties (mass, stiffness and resonant frequency) specified for the 3D printed mechanical oscillators. An adaptive mesh of hexahedral elements, generated with the commercial software Trelis 16.5 through python scripting, is adopted to speed up the computational cost of the simulations and to avoid deformed elements. The inset shows the fine mesh required to accurately model the tiny resonator structure. A 3D model of identical properties without the metastructure is used for the reference case.\\
Figure \ref{fig:area}(c) shows the shear horizontal displacement field of harmonic SH waves traveling at three different time instants ($t_1=0.0210$ s, $t_2=0.0260$ s, $t_3=0.0312$ s) at the free surface (top snapshots) and through the metasurface (bottom snapshots). As already revealed by the experimental results, when SH wavefronts impinge the resonators they are scattered and attenuated. A reverse-concavity wavefront, hallmark of a possible reflection, is less visible since the source is no longer a pulse, as in the experiment, but a continuous sinusoidal excitation. However, the wave intensity in front of the metastructure appears slightly larger of that in the reference case, meaning that part of the energy is reflected. The vertical cross-section (A-A) cuts through the middle of the model (Fig. \ref{fig:area}(d)) demonstrating the SH wave propagation with and without the metasurface. In the vicinity of the resonant frequency of the oscillators, the part of SH waves that is not reflected by the metasurface is driven under the resonators leading to a strong wave attenuation at the surface.\\
\\
\textbf{Effect of the Grading}. To investigate a possible shear wave mode conversion enhanced by a graded metasurface, as observed for Rayleigh waves \cite{AndreaAlberi}, we analyse the data recorded with metasurfaces of increasing and decreasing frequencies with respect to the propagation direction. Figure \ref{fig:f3}(a) shows an outline of the two configurations. We compare the particle velocity, measured along the middle line of the box, in the frequency domain. In the decreasing frequency case (Fig.~\ref{fig:f3}(b)), waves are progressively reflected by the mechanical oscillators and the bandgap enlarged exactly as for the "classic metawedge" case described in \cite{AndreaAlberi}. Conversely, in the increasing frequency case (Fig. \ref{fig:f3}(c)), the SH wave energy is attenuated in a frequency range, which shifts gradually towards higher values as the resonance of the mechanical oscillators increases.\\
Finally, we compare the dynamics of a metasurface embedded in the granular medium with an identical metasurface arranged on a simpler bi-layered medium, supporting the propagation of Love waves. Our aim is to further highlight how the power-law stiffness profile of the granular material changes the nature of the subsurface SH wave propagation. To this end, simulations from 300 Hz to 400 Hz imposing a prescribed surface shear horizontal displacement of ($\bar{u}_y=u_{y,0}e^{-\omega t}$) with $u_{y,0}=0.01$ m and $\omega=2\pi370$ rad/s applied as a line source are performed. We first investigate the shear wave conversion in a substrate with the simplest layering that can support the existence of Love waves \cite{Achenbach}, a thin soft layer atop a firmer half-space with homogeneous properties. Figure \ref{fig:f4}(a) depicts the computational domain, a 1 m deep bi-layered substrate 5.5 × 1 m wide, modeled using continuity conditions along $y$ and absorbing conditions on the bottom and lateral boundaries. For the surface layer we assume a thickness of 0.125 m ($H_1=0.125$ m) while for the firmer half-space 0.875 m ($H_2=0.875$ m). The substrate is characterized by the same density as the granular medium ($1600$ kg/m$^3$), while we define as compressional and shear wave velocities the speed values calculated at the middle plane of each layer: $v_{c_1}=v_{c}(H_1/2)=143.1$ m/s, $v_{s_1}=v_{s}(H_1/2)=63.8$ m/s for the surface layer and $v_{c_2}=v_{c}(H_2/2)=394.4$ m/s, $v_{s_2}=v_{s}(H_2/2)=175.8$ m/s for the half-space. 
The harmonic line source is placed 1.2 m away from the left edge. The graded metasurface of increasing frequency, obtained by varying the elastic properties of the spring elements, is made of an array of mechanical oscillators resonating from 260 to 500 Hz, and is located 1.2 m from the source. The mechanical oscillator with the resonant frequency matching the excitation frequency is located 1.7 m away from the start of the metasurface. Figure \ref{fig:f4}(b) shows the harmonic response at 370 Hz. Love waves approach the graded metasurface and travel undisturbed until reaching the resonator with matching resonant frequency. At this position, dubbed "turning point", the wavefront is mode-converted and steered into the bulk. Such Love-to-shear mode conversion \cite{LoveThrees} yields a strong wave attenuation at the surface as shown by the normalised horizontal displacement ($u_y/u_{y,max}$) measured at the surface.\\
We further investigate the wave conversion by numerically computing the metasurface dispersion curves exploiting the Bloch-wave theory. The 3D FE models of the unit cells with and without the resonators, displayed in Fig. \ref{fig:f4}(c), are obtained from those of Fig. \ref{fig:f2}(d) after replacing the power-law velocity profile of the granular medium with the bi-layered one, and modifying the elastic properties of the resonator spring elements (in order to set the resonance at 370 Hz). The dispersion relations for Love waves propagating in a bi-layer medium (green dashed curves) and interacting with a metasurface of horizontal resonators (blue curves) are depicted in Fig. \ref{fig:f4}(d). The red dashed line marks the metasurface resonance, whereas the black lines represent the dispersion curves of shear bulk waves in the upper layer and in the substrate, respectively. In agreement with the findings shown in \cite{PalermoLove}, the fundamental Love mode interacting with the metasurface resonance is split in two branches, as a result of the classic "avoided crossing" between Love waves and the resonators. 
While the lower branch approaches asymptotically the metasurface resonance, the upper one reaches for decreasing values of frequency the sound line (i.e., the dispersion curve of shear bulk waves propagating in the upper layer) at the avoided crossing region. In this area, Love waves are converted into shear bulk waves, which propagate towards the interior of the substrate.\\
We now investigate the surface-to-shear wave conversion in the granular substrate. The same frequency analysis developed for Love waves is performed after replacing the layered profile with the assumed power-law depth-dependent velocity (see Fig.~\ref{fig:f4}(e)). Figure \ref{fig:f4}(f) shows the shear horizontal displacement harmonic response at 370 Hz. At the turning point, the SH waves detach from the surface pointing towards depth but shortly after bend upward towards the surface (inset of Fig. \ref{fig:f4}(f)). This is the well-known mirage effect \cite{LiuMir} caused by the inherent power-law elastic profile of the medium. Nonetheless, wave attenuation still occurs on the surface of the graded metasurface as shown by the normalised horizontal displacement ($u_y/u_{y,max}$).\\
This latter configuration is also implemented in time domain for the experimental setup using SPECFEM3D. The computational domain, depicted in Fig.~\ref{fig:f5}(c), is the same as the one adopted in Fig. \ref{fig:area}(b). The line source, situated at a distance of 0.2 m from the left boundary, generates harmonic SH waves. We used a tapered sinusoidal signal to avoid spurious oscillations at the onset. The graded metasurface is modelled by varying the resonant frequency of the oscillators, from 280 to 490 Hz. An analogous 3D model without the metasurface is used as the reference case (see Fig. \ref{fig:f5}(a)). Time transient 3D simulations are performed for different input signal frequencies (350, 370 Hz) with and without the resonators. The shear horizontal displacement field of harmonic SH waves is shown in Fig. \ref{fig:f5}(b) for the reference case and in Fig. \ref{fig:f5}(d) for the case of the graded metasurface. The findings confirm the results of the frequency domain analysis. In the reference case, shear horizontal waves propagating in the granular half-space remain confined at the surface. In the graded metasurface, as soon as the SH wave approaches the turning point (black small square), both at 350 Hz and at 370 Hz, its trajectory is driven downward, separating from the surface to be pushed again towards the surface by the velocity gradient after a short distance. Although the surface-to-body wave-conversion phenomenon does not occur, it is important to point out how the wave intensity is attenuated after encountering the resonant units.

\section*{\large{CONCLUSIONS}}
In this work, we numerically and experimentally investigate the dynamics and the dispersion properties of shear horizontal waves localized at the surface of a granular medium and their interaction with local resonances. For the reference case (the pristine granular material without the resonators), the numerical model reveals an infinite number of SH modes in alignment with recent analytical studies \cite{Aleshin,Gusev}, however, only the first two low-order modes are detected experimentally. When SH waves approach the resonant metasurface, a hybridization between the fundamental mode and the mechanical oscillators occurs and a frequency zone of quasi-zero surface velocity arises in the spectrum from 330 to 500 Hz. This attenuation zone is confirmed by the numerical dispersion analysis, which indicates quasi-zero displacement in the surface layer including the resonant structures at the considered frequency range. As soon as the waves impinge the resonators, they are both backscattered and steered downwards, channelled between the metasurface lower boundary and the in-depth stiffer medium.\\
Subsequently we investigate the effect of a graded metasurface by spatially increasing or decreasing the resonant frequency of the oscillators. Unlike Love waves propagating in a bi-layer medium, SH waves localized at the surface of a power-law elastic profile material, are not steered into the bulk but remain confined at the surface. Although at the "turning point" the waves detach from the surface to point towards the interior of the substrate, after traveling a short distance, they are bent upwards by the depth-increasing stiffness profile.\\
Similar results may be achieved in other natural or artificial materials, which feature an inhomogeneous power-law elastic profile, as for instance, stratified soils at the seismic scale \cite{Faust}. Although large-scale experiments are needed to establish the efficiency of these resonant metastructures, the findings can be used as a starting-point for the study of more refined engineered devices for shear wave attenuation in heterogeneous substrates. More sophisticated design could, for instance, explore the adoption of non-linear kinematics and inertial amplification mechanisms \cite{Zeighami} to improve the dynamic response bandwidth of the resonators.

\newpage

\bibliographystyle{unsrt}

\hfill \break
\textbf{ACKNOWLEDGMENTS}\\
This research was partially supported by the Ambizione Fellowship PZ00P2-174009 to A.C and the ETH Research Grant (49 17-1) to E. C.\\

\newpage
\section*{}
\captionsetup[figure]{labelformat={default},labelsep=period,name={FIG.}}

\end{document}